\documentclass[fleqn,usenatbib]{mnras}

\usepackage[T1]{fontenc}

\DeclareRobustCommand{\VAN}[3]{#2}
\let\VANthebibliography\thebibliography
\def\thebibliography{\DeclareRobustCommand{\VAN}[3]{##3}\VANthebibliography}

\usepackage{graphicx}	
\usepackage{amsmath}	

\renewcommand{\vec}[1]{ {\mathbf #1} }

\newcommand{\Eq}{{Equation}}

\newcommand{\Fig}{{Figure}}

\title[Non-radial Eruption]
{Why are non-radial solar eruptions less frequent than radial ones?}

\author[Liu et al.]{Qingjun Liu,$^{1}$ Chaowei Jiang,$^{1}$ \thanks{E-mail:chaowei@hit.edu.cn (CWJ)} Xuesheng Feng,$^{1}$ Pingbing Zuo,$^{1}$ Yi Wang,$^{1}$
  \\
  $^{1}$Shenzhen Key Laboratory of Numerical Prediction for Space
  Storm, Institute of Space Science and Applied Technology,\\ Harbin
  Institute of Technology, Shenzhen 518055, China}

\date{Accepted XXX. Received YYY; in original form ZZZ}

\pubyear{2023}

\begin{document}
\label{firstpage}
\pagerange{\pageref{firstpage}--\pageref{lastpage}}
\maketitle

\begin{abstract}
Coronal mass ejections from the Sun are not always initiated along a radial trajectory; such non-radial eruptions are well known to be caused by the asymmetry of the pre-eruption magnetic configuration, which is primarily determined by the uneven distribution of magnetic flux at the photosphere. Therefore, it is naturally expected that the non-radial eruptions should be rather common, at least as frequent as radial ones, given the typically asymmetrical nature of photospheric magnetic flux. However, statistical studies have shown that only a small fraction of eruptions display non-radial behavior. Here we aim to shed light on this counterintuitive fact, based on a series of numerical simulations of eruption initiation in bipolar fields with different asymmetric flux distributions. As the asymmetry of the flux distribution increases, the eruption direction tends to deviate further away from the radial path, accompanied by a decrease in eruption intensity. In case of too strong asymmetry, no eruption is triggered, indicating that excessively inclined eruptions cannot occur. Therefore, our simulations suggest that asymmetry plays a negative role in producing eruption, potentially explaining the lesser frequency of non-radial solar eruptions compared to radial ones. With increasing asymmetry, the degree of non-potentiality the field can attain is reduced. Consequently, the intensity of the pre-eruption current sheet decreases, and reconnection becomes less efficient, resulting in weaker eruptions.
\end{abstract}





\section{Introduction}
\label{sec:intro}


Solar eruptions refer to the large-scale explosive events as powered by impulsive release of magnetic energy in the solar corona. They are often accompanied with fast outward ejections of magnetized plasma from the Sun, i.e., coronal mass ejections (CMEs). CMEs often experience deflection away from a radial direction during their propagation~\citep{wangDeflectionCoronalMass2004, shenKinematicEvolutionSlow2011, lugazNUMERICALINVESTIGATIONCORONAL2011, guiQuantitativeAnalysisCME2011, 2014EP&S...66..104G, mostlStrongCoronalChannelling2015}. Such a deflection may also occur in the very early stage when the eruption is initiated in the lower corona, 
 which can be tracked by prominence (or filament) eruptions. Observations reveal that the directions of prominence eruptions do not always adhere to a radial trajectory~\citep{2021A&A...647A..85D, 2022SoPh..297...18Z}. For a so-called non-radial eruption, the ejection apparently deviates from the radial trajectory. It is well known that the non-radial eruption is caused by the asymmetry of the pre-eruption configuration, which is mainly determined by the asymmetrical distribution of photospheric magnetic flux in the eruptive source region.
For example, \citet{2012ApJ...757..149S} studied a filament eruption that exhibited a high inclination angle of approximately $60^{\circ}$, and suggested that due to the asymmetrical distribution of photospheric flux, the confining magnetic pressure decreases more rapidly in the tangential direction than radially. Many simulations of solar eruption initiation have clearly demonstrated this. For instance, by using an asymmetric bipolar field as the initial condition, \citet{2010ApJ...708..314A} achieved an asymmetric eruption of the magnetic field with respect to the polarity inversion line (PIL). Data-constrained and data-driven simulations are also carried out to study non-radial eruptions~\citep{2013SoPh..286..453T, 2021ApJ...919...39G, 2023ApJ...956..119G,2023ApJ...947L...2Z,2021ApJ...919...39G}.


\begin{figure*}
  \centering
  \includegraphics[width=\textwidth]{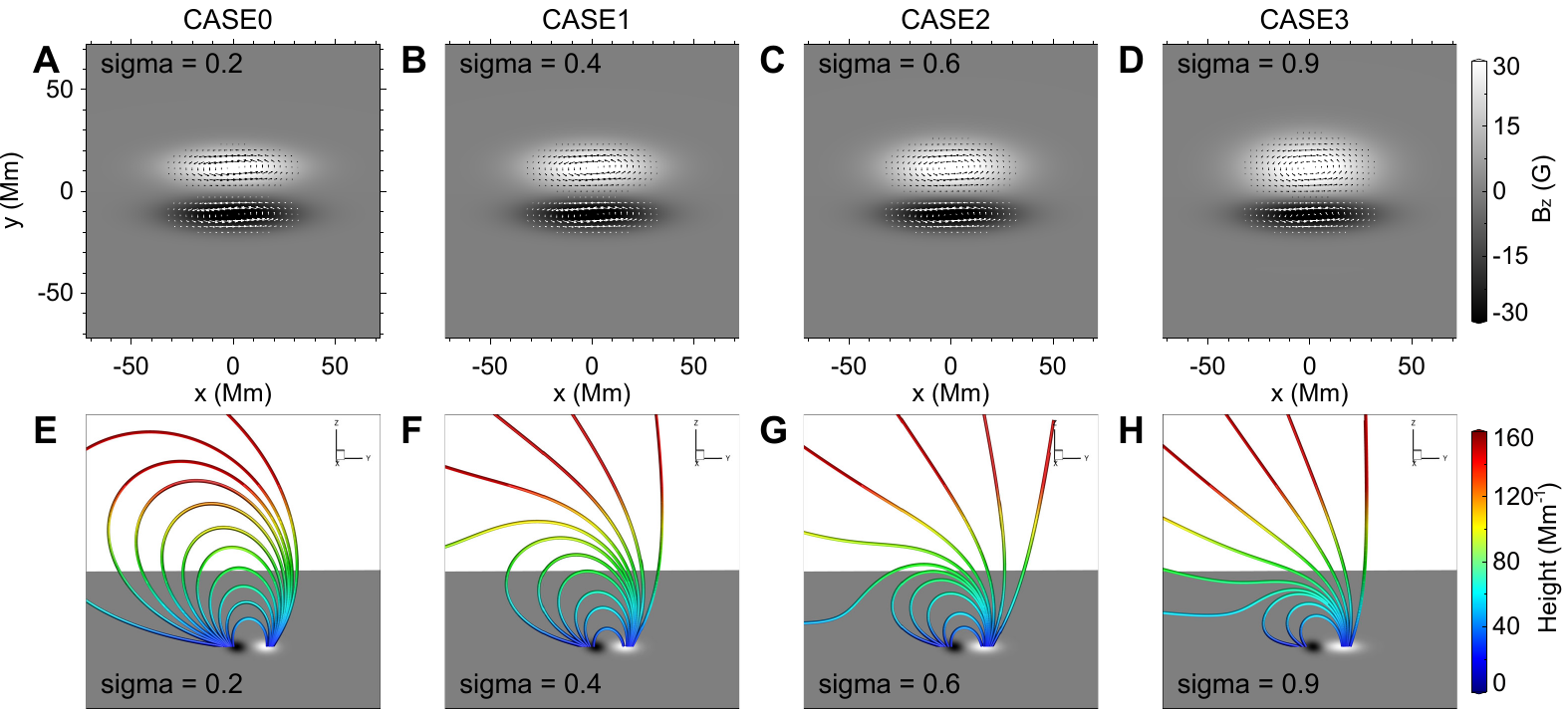}
  \caption{Magnetic flux distribution and corresponding potential field configuration in the four simulation cases with increasing degree of asymmetry. The top panels show magnetic flux distribution and surface rotation flow at the bottom surface (i.e., $z = 0$ plane). The background is color-coded by the vertical magnetic component $B_z$, and the vectors show the rotation flow. The four panels A-D refer to the four cases with $\sigma= 0.2$, $0.4$, $0.6$, and $0.9$, respectively. The bottom panels shows the initial potential magnetic field. The colored thick lines represent magnetic field lines, with the colors indicating the height. The background displays the magnetic flux distribution at the bottom surface same as panels A-D.}
  \label{f1}
\end{figure*}

Observations show that magnetic flux distribution at the photosphere is generally not symmetric with respect to the PIL. Typical bipolar active regions (ARs) consist of a leading sunspot with concentrated magnetic flux and trailing sunspots with somewhat dispersed magnetic flux. It is also often that new bipolar flux emergence into the sides of pre-existing ARs, forming complex asymmetric configurations. Therefore, it is naturally expected that the non-radial eruptions should be rather common, at least no less than the radial ones.
However, with a comprehensive survey of 904 filament eruptions observed by the Solar Dynamics Observatory in the years from 2010 to 2014, \citet{2015SoPh..290.1703M} found that only a small fraction ($\sim 20$\%, which also including the sideways event that directs almost tangentially) of the eruptions exhibit non-radial behavior. Although not explicitly discussed in the literature, such a minority of non-radial eruptions appears to be counter-intuitive ~\citep{2014EP&S...66..104G}, and the underlying reason remains unclear.

In this Letter, we attempt for the first time to offer an explanation based on a sequence of numerical experiments of eruption initiation in simple bipolar fields. These simulations are extended from our previous ones for symmetric bipolar fields~\citep{2021NatAs...5.1126J, 2022A&A...658A.174B}, which produce radial eruptions, to asymmetric cases that produce non-radial eruptions. We show that, by continuously shearing an asymmetric bipolar field, a current sheet (CS) is progressively built up above the polarity inversion line (PIL), and once reconnection starts at the CS, an eruption is initiated, which follows the same fundamental mechanism as demonstrated in the symmetric case. By comparing the different simulations of different asymmetric flux distributions, it is found that, as the degree of asymmetry increases, the eruption direction becomes more inclined away from the radial direction, while the eruption intensity becomes weaker. When the asymmetry is strong enough, no eruption can be triggered, and therefore eruption with too large an inclination angle is difficult to occur. Therefore our simulations suggest that the asymmetry plays a negative role in producing eruption, which likely explains why non-radial solar eruptions are less frequent than radial ones. We further explored the physical reason in the point of view of magnetic non-potentiality and the strength of the pre-eruption formed CS.



\begin{figure*}
  \centering	
  \includegraphics[width=\textwidth]{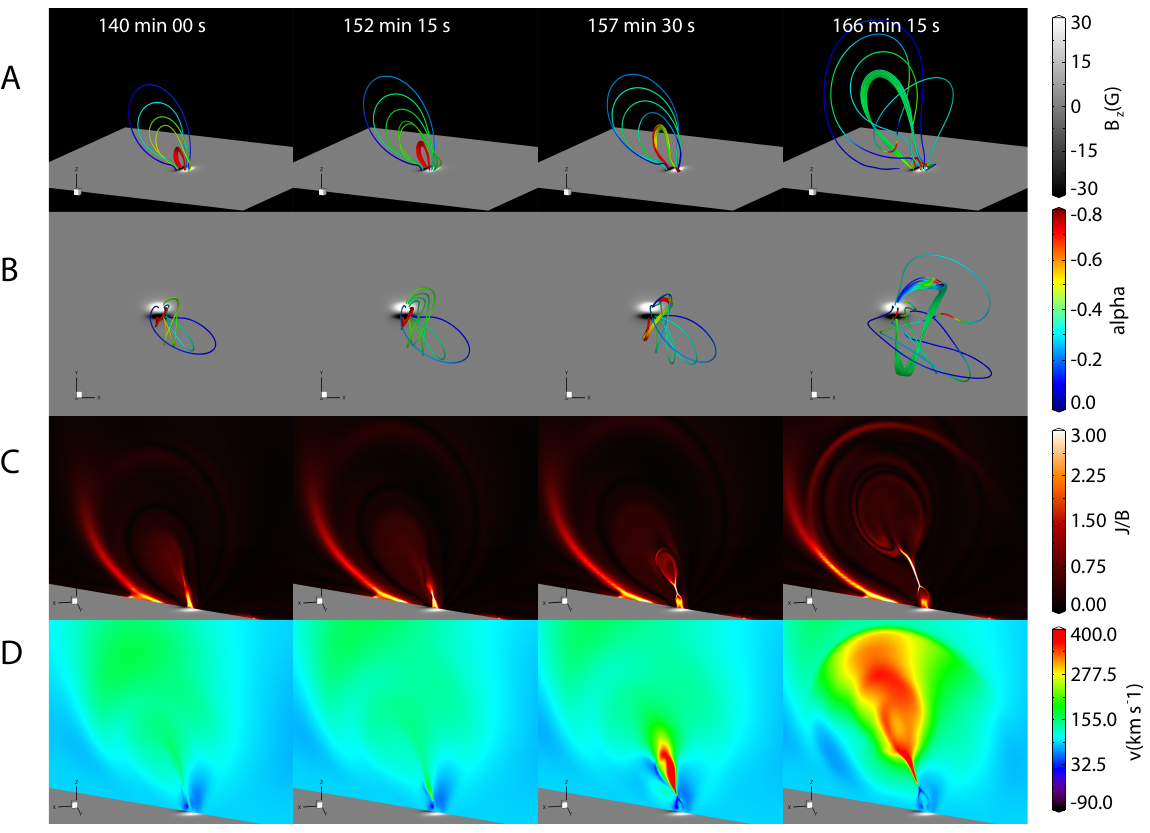}
  \centering
  \caption{Evolution of magnetic field line, current, and velocity from pre-eruption phase to eruption in the simulation of CASE~2. (A) Side view of magnetic field lines. The colored thick lines represent magnetic field lines and the colors denote the value of nonlinear force-free factor defined as $\alpha = \vec J \cdot \vec B/B^2$, which indicates the extent to which the field lines are non-potential. The background shows the magnetic flux distribution on the bottom boundary. (B) 3D \textbf{perspective} view of the same field lines shown in panel A. (C) Distribution of $J/B$, i.e., \textbf{magnitude} of current density normalized by magnetic field strength, in the vertical cross-section of volume (i.e., the $x=0$ plane). (D) Magnitudes of velocity in the same cross section as shown in panel C.}
    \label{f3}
\end{figure*}

\begin{figure*}
  \centering
  \includegraphics[width=0.8\textwidth]{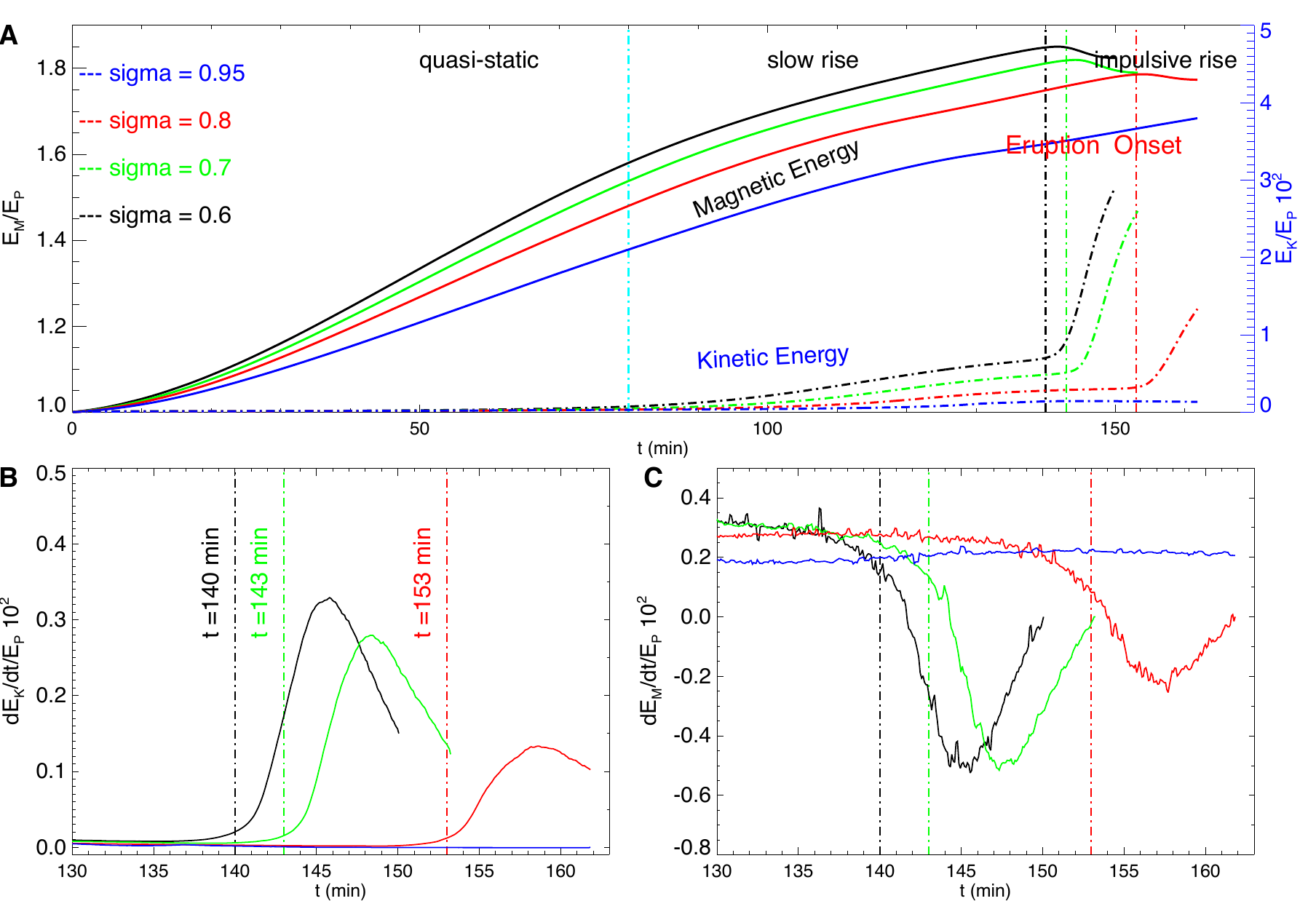}
  \caption{Evolution of different parameters of energies in all the four simulations. (A) Magnetic energy $E_M$ and kinetic energy $E_K$. The colors of the curves denote the results of the different runs; black for CASE0, green for CASE1, red for CASE2 and blue for CASE3. (B) Increasing rate of kinetic energy $d E_K/dt $ around the eruption time. (C) Releasing rate of magnetic energy $d E_M/dt$ around the eruption time. In all panels, the vertical dot-dashed lines mark the eruption onset times, which are different in the different runs, as denoted by the numbers in panel B.}
\label{f4}
\end{figure*}

\section{Experiment settings}
\label{model}
The numerical experiments are carried out by solving the full MHD equations within 3D Cartesian geometry, with the $z$ axis representing the radial direction on the Sun, and the key variables (magnetic field, plasma density and temperature) are configured to mimic typical coronal environments with low plasma $\beta$ ( $\sim 10^{-3}$) and high Alfv{\'e}n speed ($\sim 10^3$~km~s$^{-1}$).  Since the MHD code is the same one as used in \citet{2021NatAs...5.1126J}, readers are referred to that paper for more details.


The magnetic flux distribution at the bottom surface is a bipolar field~\citep{2003ApJ...585.1073A,2021NatAs...5.1126J} composed of two Gaussian functions,
\begin{equation}
B_z(x, y, 0) =B_0 e^{-x^2/\sigma_x^2} (e^{ -(y^2 -y_c^2)/ \sigma_{y_1}^2} - e^{ -(y^2 +y_c^2)/ \sigma_{y_2}^2}),
\end{equation}
where $\sigma_x $ and $ \sigma_{y_{1,2}} $ control the extents of the magnetic flux distribution in the $x$ and $y$ directions, respectively, and $y_c$ represents the distance (in $y$ direction) between the two magnetic polarities. Asymmetric distributions of the positive and negative flux can be obtained by using unequal values for $\sigma_{y_1}$ and $\sigma_{y_2}$. \Fig~\ref{f1}(A-D) shows four different cases with $ \sigma_{y_1}$ being $0.6$, $0.7$, $0.8$, and $0.95$ (normalized by a length unit of $11.52$~Mm), respectively, while the other parameters are fixed as $\sigma_x=2$, $\sigma_{y_2}=0.5$, and $y_c = 0.8$. For convenience of comparison, the degree of asymmetry, referred to as $\sigma$, is defined as
\begin{equation}\label{sigma}
  \sigma = \frac{\sigma_{y_1}-\sigma_{y_2}}{\sigma_{y_2}},
\end{equation}
and thus the four cases have asymmetry degree of $\sigma=0.2$, $0.4$, $0.6$, and $0.9$. The degree of asymmetry can also be measured by the ratio of positive to negative magnetic fluxes, $|\Phi_p/\Phi_n|$, which are $1.2$, $1.43$, $1.68$, and $2.0$, respectively, showing a linear correlation with $\sigma$.

With these different flux distributions, we use the Green's function method to obtain the initial potential field. \Fig~\ref{f1} E-H show the initial potential field of the four cases.  Due to the asymmetry, the field lines in different cases have different inclination angles, which increase with the degree of asymmetry. Also because the flux is not balanced between the two polarities, some of the field lines extend to infinity (i.e., open), which mimics solar ARs with open flux.

The computational volume spans $[-270, 270]$~Mm in the transverse directions and $[0, 540]$~Mm in the vertical direction, discretized with adaptive mesh refinement.
The highest resolution is $360$~km to capture the formation process of the CS and the subsequent reconnection.
At the bottom surface (i.e., the $z = 0$ plane), we apply slow-driving rotational flows at the footpoints of the field, which is shown in \Fig~\ref{f1}. The velocity profile is defined by
\begin{eqnarray}
	v_x = \frac{\partial \Psi(B_{z})}{\partial y} ,	v_y = -\frac{\partial \Psi(B_{z})}{\partial x}
\end{eqnarray}
where
\begin{eqnarray}
     \Psi(B_z) = v_0 B_{z}^2e^{-(B_{z}^2-B_{z,\max}^2)/B_{z,\max}^2}
\end{eqnarray}
with $B_{z,\max}$ representing the maximum value of $B_z(x,y,0)$ and $v_0$ is a constant such that the maximum speed of the flow is approximately $5$~km~s$^{-1}$. Although the speed is somewhat larger than typical photospheric flow speed ($\sim 1$~km~s$^{-1}$), the applied flow is sufficiently small when compared to the Alfv{\'e}n speed in the corona, and therefore drives the pre-eruption configuration to evolve in a quasi-static way. Moreover, this velocity profile ensures that the magnetic flux distribution remains unchanged while magnetic shear is created along the PIL. The magnetic induction equation is solved directly at the bottom boundary to update the magnetic field in a self-consistent manner as driven by the surface flow. All other surfaces are open boundaries.

We conducted four simulations, referred to as CASE 0 to CASE 3, corresponding to the four different magnetograms with increasing $\sigma$ as shown in \Fig~\ref{f1}. All the simulations begin with the potential field, along with a plasma in hydrostatic equilibrium, and is continuously driven by the flows applied at the bottom boundary.

\begin{figure*}
  \centering
  \includegraphics[width=0.8\textwidth]{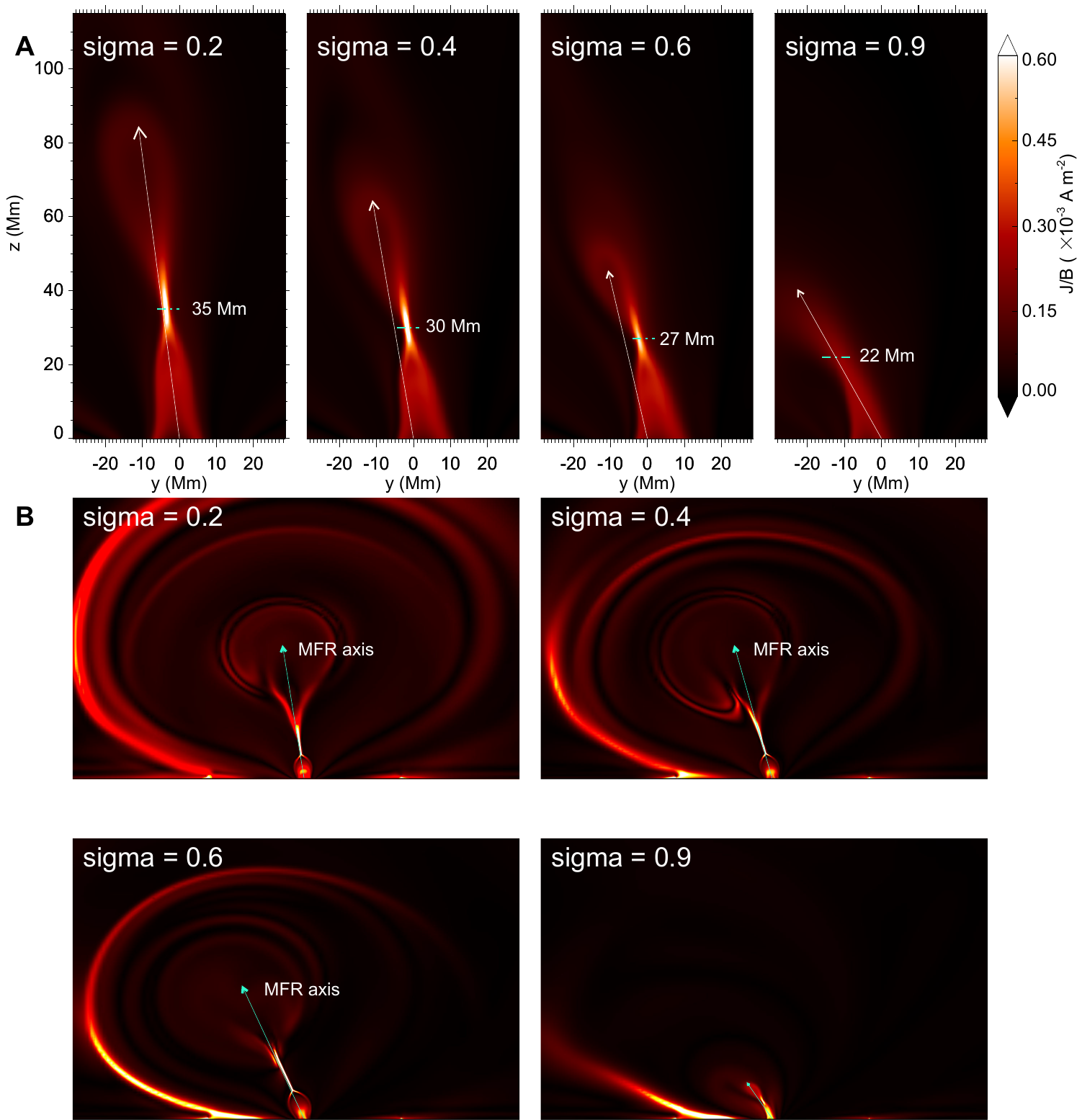}
  \caption{The directions of eruptions as shown by the CS and the MFR in the different simulations.  (A) The current density normalized by the magnetic field strength, $J/B$, in the vertical cross section of the volume at the eruption onset time. The white arrows indicate the orientation of the CS, and the cyan dashed horizontal lines denote the location of maximum current density in the CS. (B) $J/B$ in the vertical cross section at the peak time of the eruption, i.e., the time when the kinetic energy reaches its maximum. The cyan arrows denote the direction of eruption, pointing from the bottom PIL to the axis of the erupting MFR.}
    \label{f5}
\end{figure*}
	
\begin{figure*}
  \centering
  \includegraphics[width=0.8\textwidth]{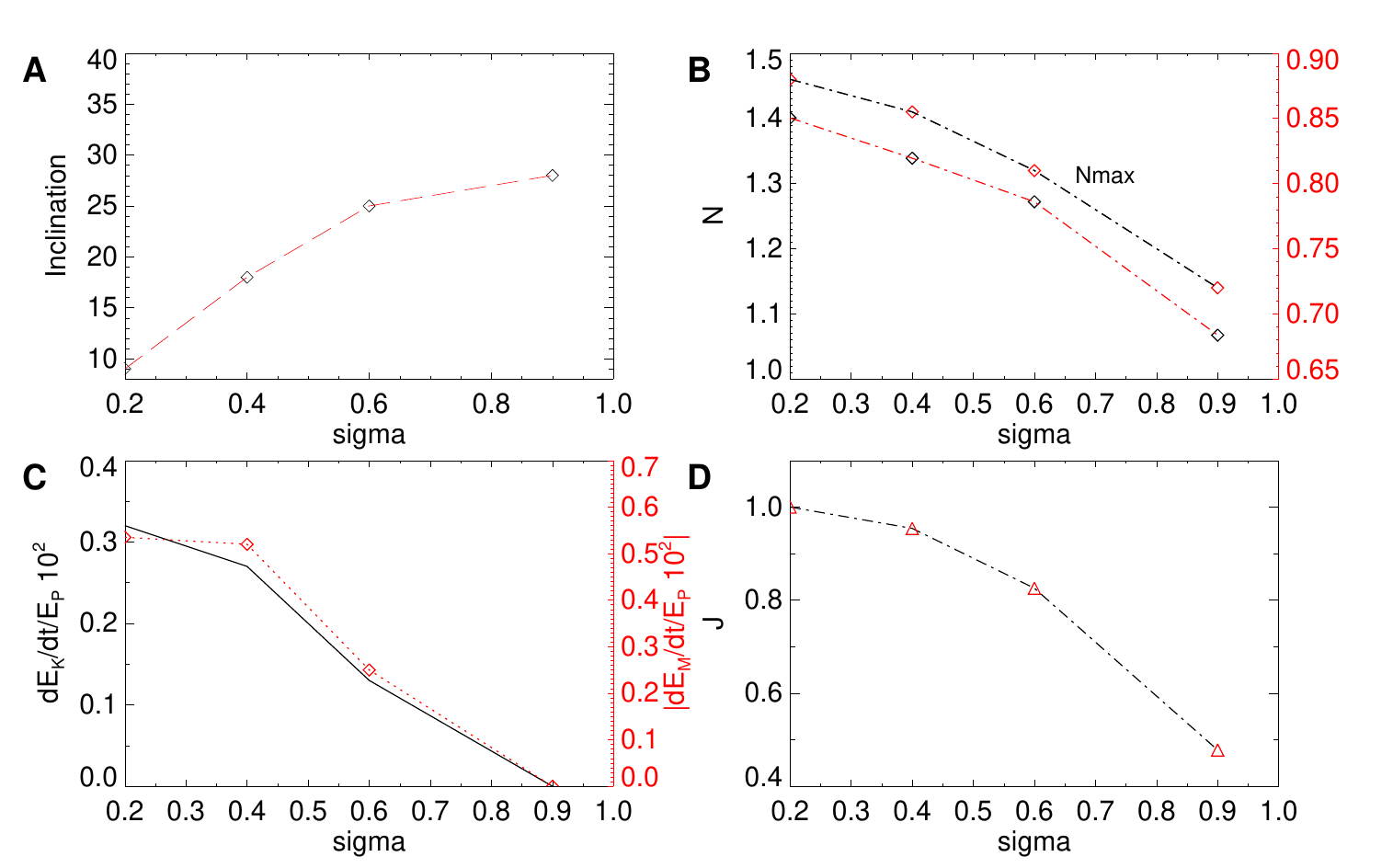}
  \caption{Dependence of different parameters of the eruption on the degree of asymmetry. (A) Inclination angle of the eruption, which is quantified by the angle formed between the MFR axis direction and the normal. (B) Non-potentiality at the eruption onset time. The black dotted line represents the non-potentiality of the corresponding open field. (C) The peak values of the kinetic energy increasing rate and the magnetic energy releasing rate. (D) the maximum current density in CS at the eruption onset time.}
  \label{f6}
\end{figure*}

\section{Results}
\label{results}
As an example, \Fig~\ref{f3} illustrates evolution of magnetic field lines, current density, and velocity in simulation of CASE 2. The evolution from pre-eruption phase to eruption follows essentially the same scenario as shown in our previous simulations for symmetric bipolar fields~\citep{2021NatAs...5.1126J, 2022A&A...658A.174B}. Prior to the eruption, the magnetic configuration consists of a sheared arcade core located above the PIL. This core contains intense currents and is surrounded by an overlying field that is nearly free of current. As the degree of magnetic shear intensifies, the magnetic pressure increases, which drives the core field to expand gradually. 
During this process, a vertical CS gradually forms above the PIL. This occurs as the distribution of currents is squeezed from a volumetric distribution into a narrow, approximately vertical layer, as depicted in \Fig~\ref{f3}C, which shows the current density normalized by the magnetic field strength (i.e., $J/B$) to emphasize thin layers with strong current. A critical transition occurs when the thickness of the CS decreases to the grid resolution and magnetic reconnection starts, triggering an eruption. The evidence lies in the topology change of the core magnetic field: it shifts from a sheared arcade to a flux rope, as illustrated in \Fig~\ref{f3} A and B. Additionally, reconnection flows are observed during this process, as the velocity distribution in \Fig~\ref{f3}D clearly shows that the outflows on top and bottom sides of the reconnection site.

\Fig~\ref{f4} shows the temporal evolution of magnetic and kinetic energies (and their changing rate) of all the cases from CASE 0 to 3 (note that the energies are normalized by the potential field energy $E_p$ corresponding to each magnetogram). In the first three cases, there is a typical slow storage of magnetic free energy leading to its fast release in an eruption. The transition from the pre-eruption phase to eruption onset is clearly manifested in the evolution of kinetic energy. Notably, there is a sharp transition at $140$~min for CASE~0, $143$~min for CASE~1, and $153$~min for CASE~2.	 In details, the evolution can be divided into three stages. The first stage, known as the quasi-static phase (from the beginning to around $t = 70$~min), is dominated by the shearing flow at the bottom. During this phase, the core field expands at a speed close to that of the bottom driving speed. Simultaneously, the magnetic energy injected from the bottom boundary is stored in the coronal magnetic field with almost no loss. Therefore, the magnetic energy increases monotonically and linearly, while the kinetic energy remains almost unchanged, being three orders of magnitude smaller than the magnetic energy. The second stage is a slow rise phase (from around $t = 70$~min to the onset time of eruptions in the different cases). In this phase the magnetic energy continues to increase, but at a slower rate than in the quasi-static phase. The kinetic energy begins to deviate evidently from the initial value and slowly increases, because the field expands faster than during the quasi-static phase. The final stage is the eruption phase, which begins when fast reconnection is triggered at the central CS. The magnetic energy rapidly releases about $5\%$ during this process, even though the boundary driving still injects magnetic energy into the volume. The kinetic energy sharply increases by about $0.03~E_p$ within $10$ minutes.		

We systematically examine the directions of eruption across the four cases. As shown in \Fig~\ref{f5}, we measure the inclination angle of the eruption direction from the $z$ axis in the $x=0$ plane, which intersects perpendicularly the axis of the erupting MFR. The direction of eruption is defined as the line pointing from the origin point to the rope's axis (as depicted in \Fig~\ref{f5}B). The result is shown in \Fig~\ref{f6}A. Notably, this inclination angle increases from approximately $10^{\circ}$ to around $30^{\circ}$ as $\sigma$ varies from $0.2$ to $0.9$, which clearly shows that the eruption direction exhibits a consistent variation with the degree of asymmetry of flux distribution.


The intensity of eruption also varies in the different cases. The eruption intensity can be measured by the peak value of kinetic energy increasing rate, or the peak value of the magnetic energy release rate during the eruption (as seen in \Fig~\ref{f4}B). \Fig~\ref{f6}C shows the relationship between asymmetry and eruption intensity. Notably, the eruption intensity demonstrates a negative correlation with asymmetry, and it decreases to almost $0$, i.e., no eruption occurs, when $\sigma$ reaches $0.9$. This result suggests that for a simple bipolar configuration, the asymmetry does not favor for eruption, and no eruption can be triggered when the asymmetry is too strong. This might explain why non-radial eruptions with large inclination angle, due to the asymmetry of the flux distribution, is less observed than the radial eruptions.

In the experiments with a range of different symmetric bipolar fields~\citep{2022A&A...658A.174B}, it has been shown that the intensity of eruptions hinges upon the strength of the pre-eruption formed CS, which is closely related to the degree of the field non-potentiality. Specifically, the higher the non-potentiality can be reached, the higher the current density in the CS is; Consequently the more efficient the reconnection can be, and the stronger the resulted eruption is. Moreover, for a given magnetic flux distribution, the upper limit of non-potentiality achievable is determined by the open field~\citep{alyHowMuchEnergy1991, sturrockMaximumEnergySemiinfinite1991}. Although such an upper limit cannot be reached in numerical simulation, it is systematically correlated with the actual obtained value. Here we apply the same analysis as shown in~\citet{2022A&A...658A.174B}; The field non-potentiality, $N = E_{f}/E_{p}$, is defined as the ratio of free magnetic energy $E_f$ to potential field energy $E_p$. Its upper limit is $N_{\rm max} = (E_{o} - E_{p})/E_{p}$, where $E_o$ is the open field energy\footnote{The formula of computing the open field energy can be found in \Eq~(4) of \citet{2022A&A...658A.174B}}. Note that $N_{\rm max}$ is totally determined by the magnetic flux distribution.

\Fig~\ref{f6}B shows the relationship of $N$ and $N_{\rm max}$ at the eruption onset with the degree of asymmetry. As can be seen, the overall decrease in non-potentiality (both $N$ and $N_{\rm max}$) aligns with the increase in asymmetry degree.  The decrease of $N_{\rm max}$ can be understood from the point of view of the open field configuration; if the field asymmetry degree is larger, more flux is open initially in the potential field, and thus the non-potentiality it can attain is reduced.

\Fig~\ref{f6}D depicts the maximum current density in the CS at the eruption onset time, which also decreases as the field asymmetry intensifies. Therefore the variation of intensity of eruptions in the asymmetry fields also follows the same rule as found in the symmetric fields.



\section{Summary}
\label{concl}
By extending our previous numerical simulations~\citep{2021NatAs...5.1126J, 2022A&A...658A.174B}, which established a fundamental mechanism of solar eruption initiation, from symmetric bipolar fields to asymmetric cases, we surveyed how the eruption initiation process is influenced by the degree of the asymmetry, which can be simply measured by the magnetic flux ratio of the two polarities. The simulations show that the asymmetric bipolar fields produce non-radial eruptions, and their inclination angle from radial direction increases as the degree of asymmetry increases. Importantly, the intensity of eruption exhibits a strong negative correlation with the degree of asymmetry (and thus also the inclination angle), and no eruption can be triggered when the asymmetry is strong enough or the inclination angle is large enough. For the first time, this provides an possible explanation for the observation fact that non-radial eruptions are less frequent than radial ones. We further explored the physical reason why the asymmetry does not favor for eruption. When the degree of asymmetry of the bipolar field increases, the degree of non-potentiality the field can attain is reduced. Consequently, there is a decline in the intensity of the pre-eruption formed CS, and the reconnection becomes less efficient, which in turn weakens the strength of its resulted eruption. Finally, we note that our simulations are only apply to the bipolar fields. The direction of eruption might also be caused by other factors, such as complex magnetic topology with multipolar configurations, which needs to be investigated in future study.



\section*{Acknowledgements}
This work is jointly supported by National Natural Science
Foundation of China (NSFC 42174200), Shenzhen Science and Technology
Program (Grant No. RCJC20210609104422048), Shenzhen Key Laboratory Launching Project
(No. ZDSYS20210702140800001), Guangdong Basic and Applied Basic
Research Foundation (2023B1515040021) and the Fundamental Research Funds for the Central Universities (Grant No. HIT.OCEF.2023047). 

\section*{Data Availability}
All the data generated for this paper are available from the authors
upon request.













\bsp	
\label{lastpage}
\end{document}